\documentclass{aastex}
\usepackage{emulateapj5}

\newcommand{\Halpha}{H$\alpha$}

\shorttitle{SFR dependent obscuration in galaxies}
\shortauthors{Afonso et al.}

\begin{document}

\title{Dependence of dust obscuration on star formation rates in galaxies}

\author{J. Afonso\altaffilmark{1,2,3}, A. Hopkins\altaffilmark{4,5}, B. Mobasher\altaffilmark{6,7}, and C. Almeida\altaffilmark{1,8}}

\altaffiltext{1}{CAAUL, Observat\'orio Astron\'omico de Lisboa, Tapada da Ajuda, 1349-018 Lisboa}
\altaffiltext{2}{Onsala Space Observatory, S-43992 Onsala, Sweden}
\altaffiltext{3}{email: jafonso@oal.ul.pt}
\altaffiltext{4}{Department of Physics and Astronomy, University of Pittsburgh, 3941 O'Hara Street, Pittsburgh, PA 15260, USA; ahopkins@phyast.pitt.edu}
\altaffiltext{5}{Hubble Fellow}
\altaffiltext{6}{Space Telescope Science Institute, 
3700 San Martin Drive, Baltimore MD 21218, USA; b.mobasher@stsci.edu}
\altaffiltext{7}{Also affiliated to the Space Sciences 
Department of the European Space Agency}
\altaffiltext{8}{email: cesario@oal.ul.pt}

\begin{abstract}
Many investigations of star formation rates (SFRs) in galaxies have
explored details of dust obscuration, with a number of recent analyses
suggesting that obscuration appears to increase in systems with high
rates of star formation. To date these analyses have been primarily based on
nearby ($z \le 0.03$) or UV selected samples. Using 1.4\,GHz imaging
and optical spectroscopic data from the {\em Phoenix Deep Survey}, 
the SFR-dependent obscuration is explored.
The use of a radio selected sample shows that previous studies exploring
SFR-dependent obscurations have been biased against obscured 
galaxies. The observed relation between obscuration and SFR is found to
be unsuitable to be used as an obscuration measure for individual
galaxies. Nevertheless, it is shown to be successful as a first order 
correction for large samples of galaxies where no other measure of 
obscuration is available, out to intermediate redshifts ($z\approx 0.8$). 
\end{abstract}


\section{INTRODUCTION}

Dust obscuration is currently recognized as one of the most serious
sources of uncertainty in studies of galaxy evolution. With the recent 
results of far-infrared (FIR) and sub-mm observations revealing an ever 
increasing number of dusty star forming galaxies 
\citep[e.g.,][]{Genzel00,Ivison00,Smail02}, the need
to unify measures of star formation rate (SFR) from independent 
indicators at different wavelengths (e.g., UV, H$\alpha$, FIR, 
radio continuum) is as pressing as ever.

A relatively simple prescription for dust extinction
correction to SFR has been suggested by \citet{Hopkins01} and 
\citet{Sullivan01}, by assuming a luminosity- (or SFR) dependent 
obscuration. This was shown to provide a 
good first order correction to optically derived SFRs, while smaller
differences still remain between different SFR indicators that are 
likely to be related to different star formation histories and/or 
extinction properties \citep{Sullivan01}. 

The above studies were based on the comparison of different SFR  
indicators for samples of relatively low redshift galaxies, selected 
at optical or UV wavelengths, which are prone to dust induced biases. 
Furthermore, at intermediate redshifts
where the H$\alpha$ line falls out of the optical window and the 
[O{\sc ii}]$\lambda$3727 line can instead be used to measure 
the SFR ($0.3<z<0.8$), a 
direct comparison is more difficult \citep[see][]{Cardiel03}. 
In this paper we explore the 
validity of the luminosity- (or SFR) dependent obscuration  
at both low ($z\lesssim 0.3$) and intermediate ($0.3<z<0.8$)  
redshifts, using a sample of star-forming galaxies selected at 
radio wavelengths (which are insensitive to dust obscuration) 
with spectroscopic information. 

Throughout this paper we adopt 
$H_0 = 70\,$ km\,s$^{-1}$\,Mpc$^{-1}$, $\Omega_M = 0.3$, and $\Omega_\Lambda = 0.7$.

\section{OBSERVATIONS}

The {\em Phoenix Deep Survey\/} (PDS) includes a 1.4\,GHz survey made 
using
the Australia Telescope Compact Array (ATCA) covering a field a little
over 4.5 square degrees, selected to lie in a region of low optical
obscuration and devoid of bright radio sources
\citep[][and references therein]{Hopkins03a}.
Optical imaging at the Anglo-Australian Telescope (AAT) has produced
an optical catalogue probing to $R=22.5$ for the whole field, and
2dF multi-object spectroscopy from the AAT provides spectra for many
of the optically identified radio sources \citep{Georgakakis99, Afonso02}.
The spectra were taken using the low resolution gratings 270R, 316R and 300B.
Most objects were observed through only one of the red or blue gratings,
with a small number being observed through both. The 270R and 316R gratings
provide a wavelength coverage of $5000 \lesssim \lambda \lesssim 8500\,$\AA,
and the 300B gives $4000 \lesssim \lambda \lesssim 7000\,$\AA. 
Redshifts were determined by visual inspection of the spectra, and line
parameters were measured through Gaussian fitting using the {\sc splot}
package in IRAF. There are currently a total of 445 galaxies with measured
spectra, of which 138 were securely classified as star forming using 
optical emission line diagnostic diagrams. A more detailed account of
the spectroscopic data reduction and classification can be found elsewhere
\citep{Georgakakis99, Afonso02}.

\section{Star formation rates in the faint radio population} \label{sect:sfr}

\begin{figure}
\epsscale{1.0}
\plotone{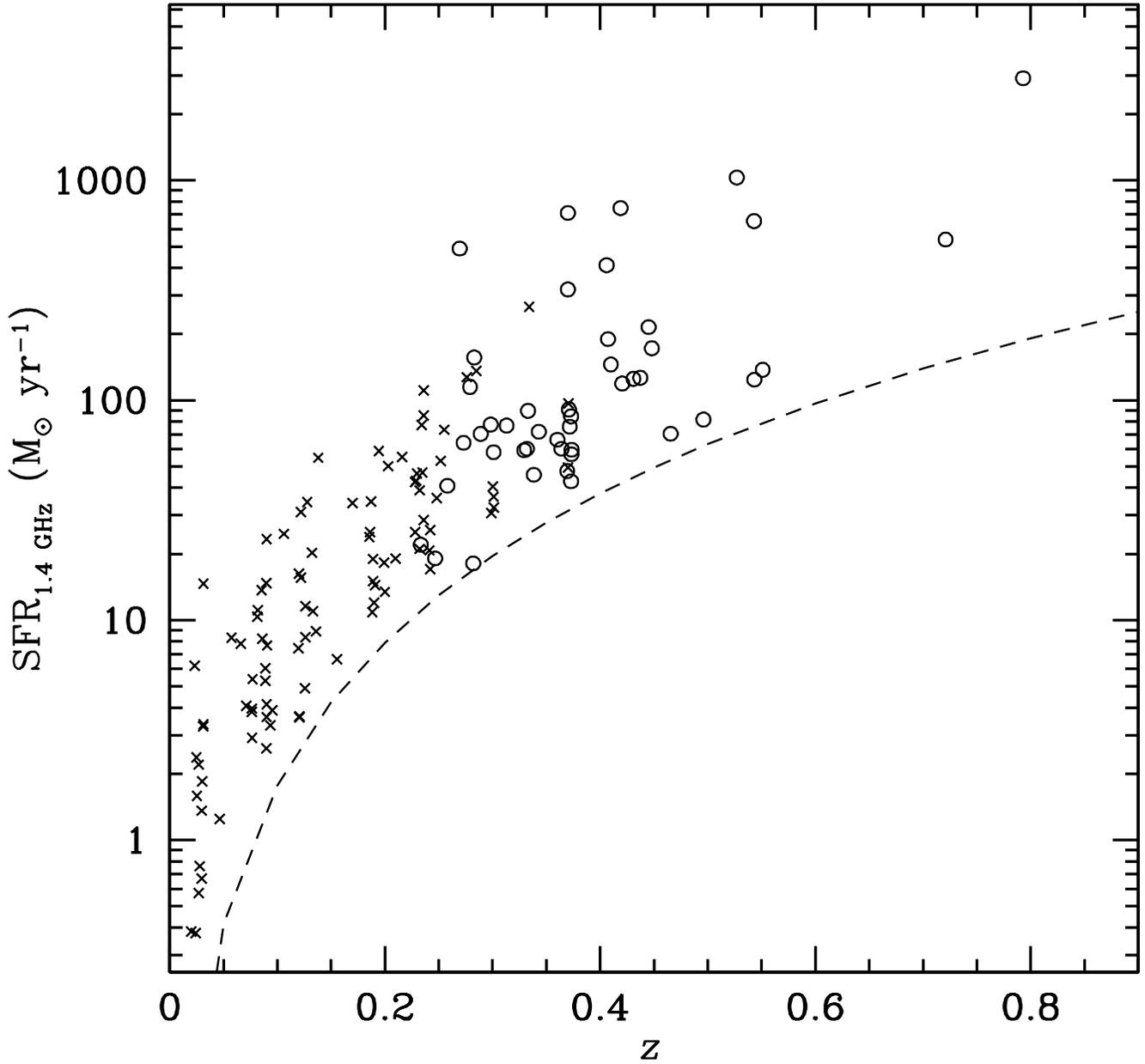}
\caption{SFRs for star-forming galaxies in
the Phoenix Deep Survey, as estimated from their radio luminosity.
Crosses indicate galaxies with an H$\alpha$ measurement suitable for a SFR 
estimate, while circles denote those where only [O{\sc ii}]$\lambda$3727 
is measured. The dashed line indicates
the 60\,$\mu$Jy detection limit of the radio observations.
\label{fig:sfr14z}}
\end{figure}

The origin of 1.4 GHz emission in star-forming galaxies is primarily
thought to be synchrotron radiation from relativistic electrons, accelerated
by the shocks from supernova ejecta. The insensitivity of radio wavelengths to
dust obscuration, makes radio emission a particularly attractive way of
estimating SFRs in star-forming galaxies. The relation between SFR
and 1.4\,GHz luminosity can be written as 

\begin{equation}
{\rm SFR}_{\rm 1.4\,GHz} = \frac{L_{1.4\,{\rm GHz}}}{8.4\times10^{20}\,{\rm W~Hz^{-1}}}~{\rm M}_{\sun}\,{\rm yr}^{-1},
\label{eqn:sfr14}
\end{equation} 

\noindent for a Salpeter IMF with stellar masses between $0.1-100$\,M$_{\sun}$
\citep{Haarsma00}. Using this relation, and calculating 
$L_{1.4\,{\rm GHz}}$ assuming a radio
spectral index $\alpha$ of 0.8 (S$_{\nu}\!\propto\!\nu^{-\alpha}$), 
characteristic of synchrotron radiation, the
SFRs for the star-forming galaxies in the Phoenix spectroscopic sample 
were calculated and are presented in Figure~\ref{fig:sfr14z}. 
The depth and area of the radio survey and the spectroscopic follow-up 
allow sampling star-formation rates from below one to a 
few hundred solar masses per year.

Of particular interest is the comparison between the SFRs as given 
by the radio and the optical line emission. To calculate 
the SFR from \Halpha, the conversion indicated by \citet{Kennicutt98}, for 
an IMF and mass limits as for equation~(\ref{eqn:sfr14}), is used

\begin{equation}
{\rm SFR}_{\rm H\alpha} = \frac{L_{\rm H\alpha}}{1.27 \times 10^{34}\,{\rm W}}~{\rm M}_{\sun}\,{\rm yr}^{-1}.
\label{eqn:sfrha}
\end{equation} 

The choice of a different IMF would change equations~(\ref{eqn:sfr14}) and 
(\ref{eqn:sfrha}) by a similar factor (which for the most commonly used 
IMFs would be $\sim$2-4), since both SFR indicators are sensitive to stars 
in roughly the same mass range (${\rm M \gtrsim 8\,M_{\sun}}$). 
Hence, a comparison
between the SFRs derived from radio and H$\alpha$ emission will be 
insensitive to the particular IMF shape choosen.

For galaxies at higher redshift ($z\,\gtrsim\,0.3$), where \Halpha\ could not 
be measured, one can use the [O{\sc ii}]$\lambda$3727 flux. 
This method relies on equation~(\ref{eqn:sfrha}) and the observed 
ratio between [O{\sc ii}] and \Halpha. A value of 
$F_{\rm [OII]}=0.45 \times F_{\rm H\alpha}$ \citep{Kennicutt98} is 
commonly used. However, using the first data release of the 
Sloan Digital Sky Survey \citep{Abazajian03}, \citet{Hopkins03b} 
observes a median relation for radio-detected star-forming galaxies of 
$F_{\rm [OII]}=0.23 \times F_{\rm H\alpha}$ (with a scatter of around 0.1), 
finding the difference to be due to the incompleteness of previous samples. 
The range of optical luminosities sampled by \citet{Hopkins03b} is 
similar to the one 
in the present Phoenix spectroscopic sample ($M_{R}\sim 20 - 23.5$) and 
is not enough to reveal the optical luminosity dependence in the 
[O{\sc ii}]/Ha ratio observed by \citet{Jansen01}, that produces lower 
[O{\sc ii}]/Ha ratios for galaxies with higher optical luminosities.
Also, the range of extinctions, as given by the Balmer decrement, sampled 
by the \citet{Hopkins03b} sample (\Halpha/H$\beta$, between 3 and 12) 
is similar to that of the present work (as will be seen below). 
We thus adopt the determination of \citet{Hopkins03b} to convert our
measured [O{\sc ii}] luminosities to \Halpha\ values, using 
equation~(\ref{eqn:sfrha}) to obtain the corresponding SFR estimate.

Figure~\ref{fig:sfr14sfrha} compares the SFRs derived from the radio 
luminosity and line emission. Although a correlation exists, as expected, 
the SFRs derived from 1.4\,GHz are in general higher that those calculated 
from nebular lines, especially for higher luminosities (or SFRs). 
This effect, seen previously in several studies 
\citep{Cram98,Hopkins01,Sullivan01}, is attributed
to dust obscuration, which affects the optical line emission.
Furthermore, one can conceive an amount of obscuration which increases
with the SFR (a SFR-dependent dust obscuration). This has been explored with
considerable success for nearby optical \citep{Hopkins01} and UV selected 
galaxies \citep{Sullivan01}. 

\section{SFR-dependent dust obscuration}

\begin{figure}
\epsscale{1.0}
\plotone{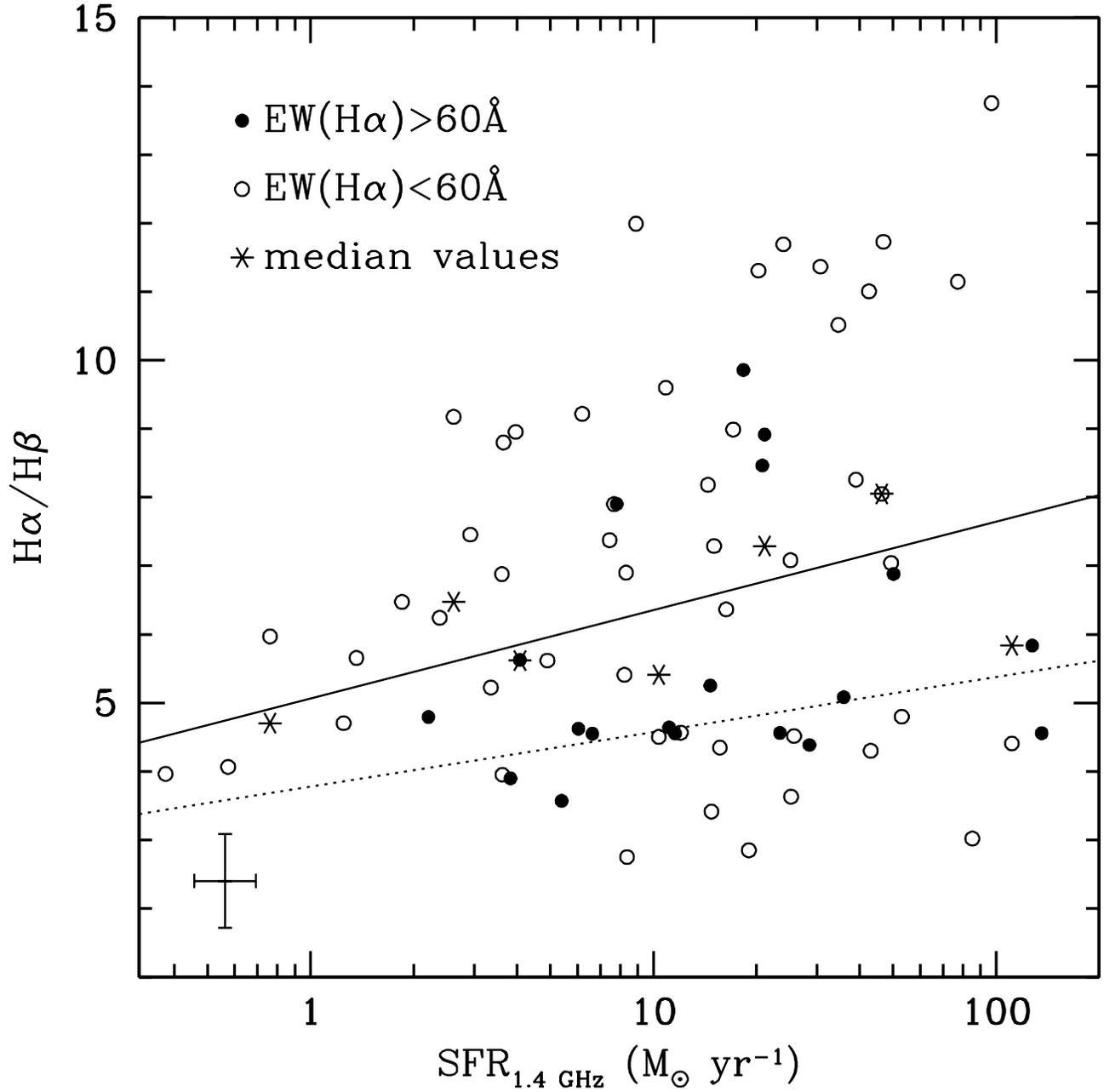}
\caption{Balmer decrements, corrected for stellar absorption,  
as a function of SFR derived from radio luminosities. 
The solid line indicates the best fit to the median values (asterisks)
in bins of $\log$(SFR), while the dotted line shows the relationship derived
in \citet{Hopkins01} for a sample of local galaxies. 
The error bar in the bottom left corner shows the median errors for the 
Balmer decrement and radio flux measurements.
\label{fig:balmer}}
\end{figure}

For a subset of the Phoenix sample described above, an estimate of the 
extinction can be made from the observed Balmer decrement 
(\Halpha/H$\beta$). Stellar absorption of the Balmer 
lines was corrected by assuming an average 
value of 2\,\AA\ for the equivalent width (EW) of the H$\beta$ 
absorption in star-forming galaxies \citep{Tresse96,Georgakakis99}, with
a similar value (2.1\,\AA) being used for the \Halpha\ line
\citep[equation (2) of][]{Miller02}.
Figure~\ref{fig:balmer} shows the resulting Balmer decrements, 
corrected for stellar absorption, as a function of SFR, derived 
from the radio luminosity using equation~(\ref{eqn:sfr14}). 
Unlike previous studies \citep{Hopkins01,Sullivan01}, a tight 
correlation is not observed here. Rather, a trend for a broader 
range of obscurations for higher SFR systems is observed. 
This behaviour seems to be due to different selection 
criteria and small number statistics, as we now explain. 

The sample used by 
\citet{Hopkins01} to derive the relation 
between SFR and obscuration includes only nearby 
($z\le 0.03$) 
galaxies with EW(\Halpha) larger than 30\,\AA. 
While no clear trend is seen in the present sample when restricted
to this EW value, the higher limit of 60\,\AA\ (filled circles in 
Figure~\ref{fig:balmer}) does suggest a closer
match to the observations in \citet{Hopkins01}. 

On the other hand, \citet{Buat02} also observed a dual behaviour: 
while nearby
star forming galaxies behave similarly to what is present 
in Figure~\ref{fig:balmer}, a sample of IUE galaxies shows a 
much tighter correlation, as that observed for the UV-selected sample
of \citet{Sullivan01}. This suggests that UV-selection results in
some kind of bias that is avoided with the present sample.
Completeness in Figure~\ref{fig:balmer} is not easy to quantify, 
given the several selection criteria 
(initially the radio flux limit, followed by the optical identification 
and 2dF spectroscopy, which imposes a practical limit of $R \sim 20$).
However, it is possible to try to understand a possible bias in a 
magnitude-limited UV selected sample, and at the same time to
evaluate the improvement of the present work, as we now show.

\begin{figure}
\epsscale{1.0}
\plotone{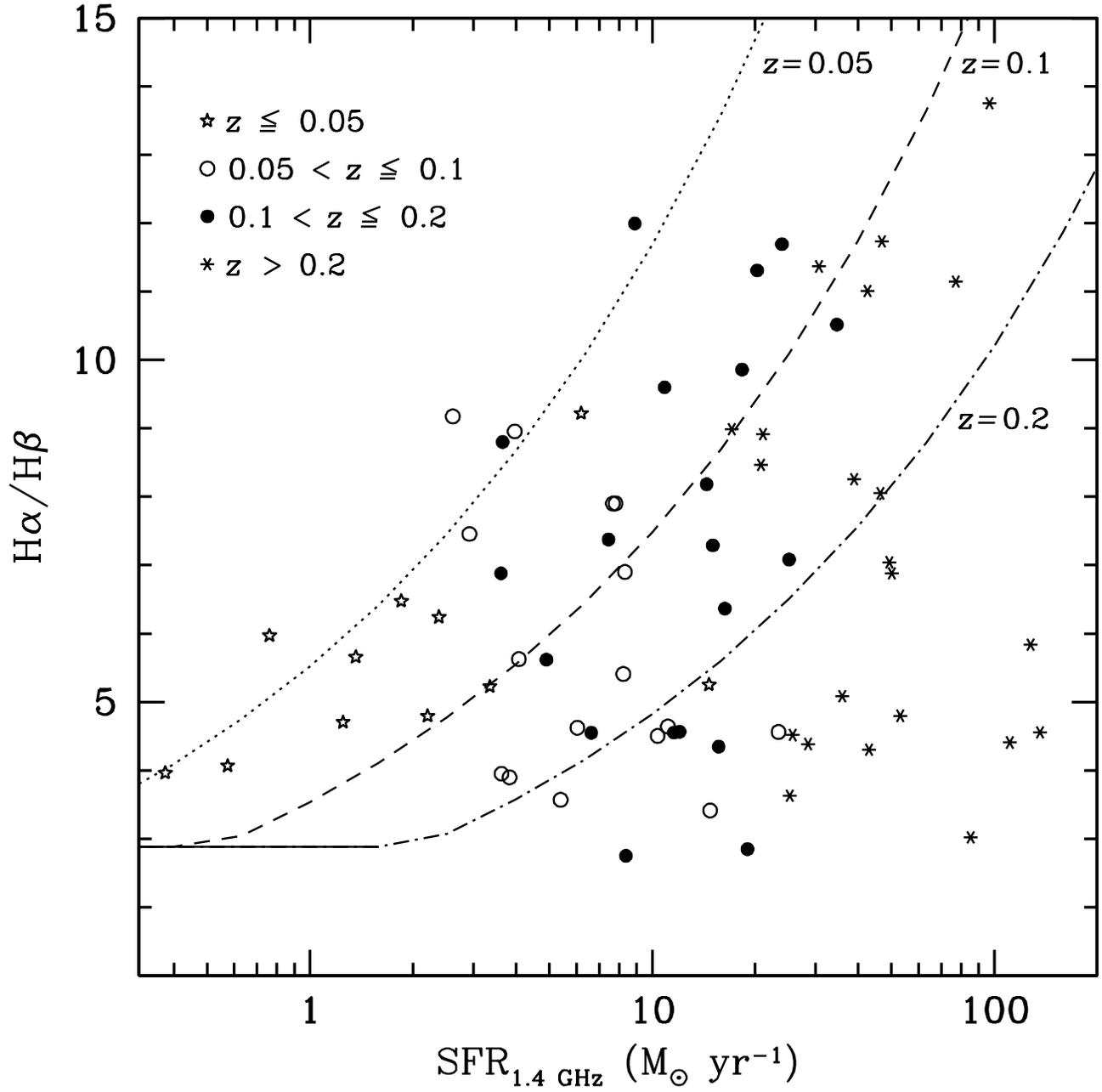}
\caption{Data as in Fig.~\ref{fig:balmer}, marked according to redshift. 
The lines indicate the maximum Balmer decrement detectable for a given
SFR at redshifts of 0.05, 0.1 and 0.2, in a UV survey with a limiting
magnitude of $m_{\rm UV}=18.5$. It is clear that most of the galaxies 
detected here displaying the highest Balmer decrements, 
would not be detected in such a UV survey.
\label{fig:balmerarea}}
\end{figure}

The tight relation between dust-free UV emission and SFR can 
be used to evaluate
which regions of Figure~\ref{fig:balmer} are not accessible to a 
magnitude-limited UV study. Assuming a limiting 
magnitude $m_{\rm UV}=18.5$  \citep[as in][]{Sullivan01}, an intrinsic SFR 
(ie, $L_{\rm UV}$ before obscuration) and redshift will define the maximum 
value of the Balmer decrement that still allows a detection. 
Figure~\ref{fig:balmerarea} shows the present sample, separated 
according to redshift,
overlaid with the maximum detectable Balmer decrement at 
$z=0.05$ (dotted line), 0.1 (dashed line) and 0.2 (dot-dashed line). 
The conversion between SFR and $L_{\rm UV}$ uses the calibration from 
\citet{Kennicutt98}, while the extinction at 2000\,\AA\ is derived from the 
Balmer decrement using the procedures of \citet{Calzetti00}. An estimate of 
the $K$-correction is obtained using an average colour of
$m_{\rm UV}-b=-1.5$ \citep{Milliard92}. 
It is clear that the present sample represents a significant improvement 
for $z>0.1$. In particular, many of the galaxies showing high 
Balmer decrement values in the present study would not be detected in a
UV survey limited to $m_{\rm UV}=18.5$. Sample selection thus seem to be a 
major source of bias when trying to investigate the correlation
between dust obscuration and SFRs. 

Given the large scatter present in Figure~\ref{fig:balmer}, a
SFR-dependent reddening correction is obviously unsuitable for
application in galaxies where a direct estimate of obscuration exists.
However, a trend for higher average Balmer decrement (and greater 
distribution width) with increasing SFR seems to exist.
This can still be useful as a preliminary dust obscuration 
estimate for large samples of galaxies where no other measure of 
obscuration is available.
Although in practice the form of the derived relation may be comparable 
to the ones in \citet{Hopkins01} and \citet{Sullivan01}, here we recognize
that there is no tight correlation between obscuration and SFR, but
an average obscuration may still be defined for any given SFR. 
As can be seen in Figure~\ref{fig:balmer} the resulting correction 
will be affected by large uncertainties for individual galaxies, 
especially at large SFRs.

\begin{figure}
\epsscale{1.0}
\plotone{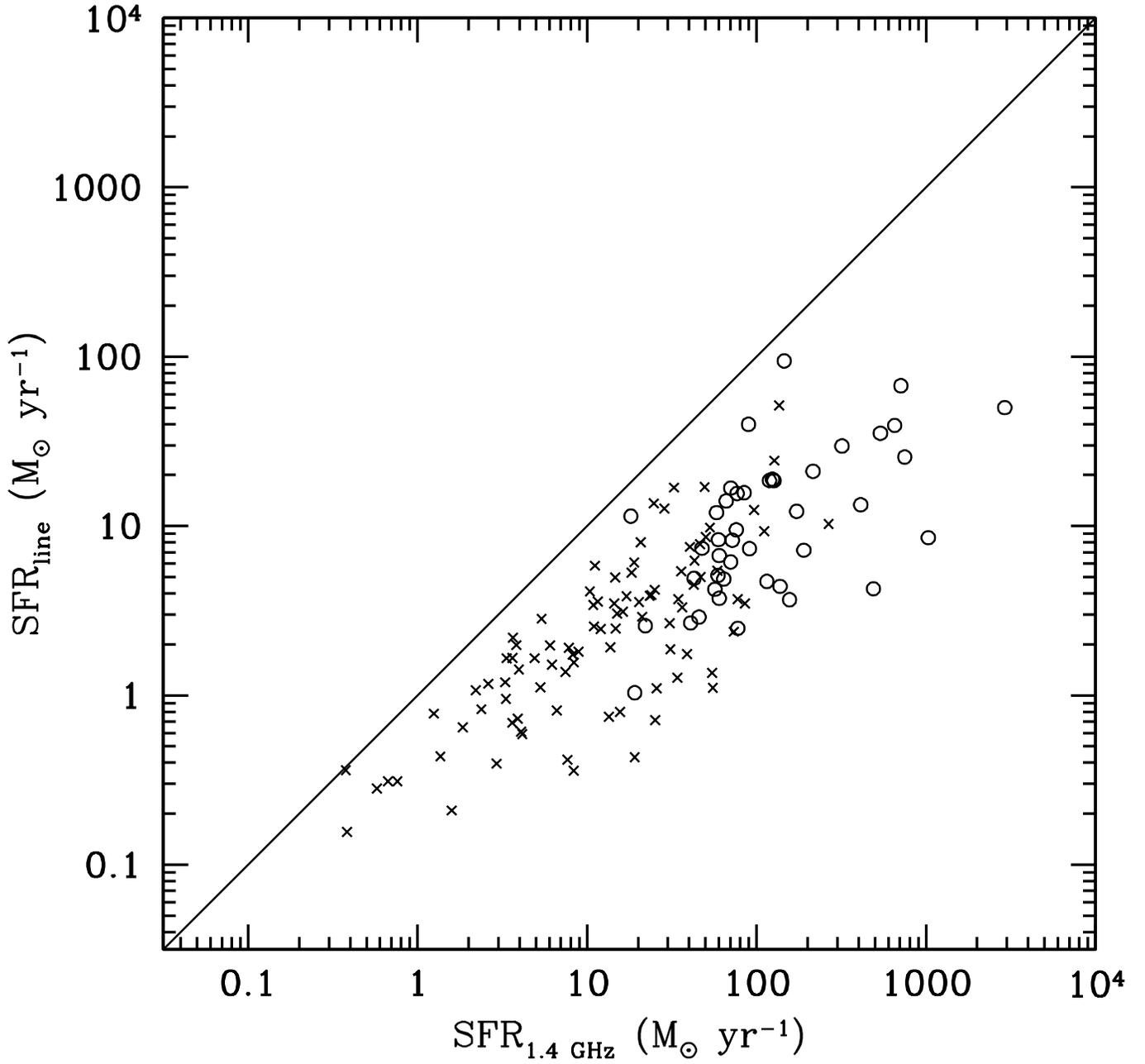}
\caption{Comparison between the SFRs derived from 1.4\,GHz and from 
\Halpha\ or [O{\sc ii}] luminosities. The crosses represent galaxies with 
direct \Halpha\ measurement, circles denote galaxies at higher redshift, 
where [O{\sc ii}] was used as a proxy for the \Halpha\ line intensity.
\label{fig:sfr14sfrha}}
\end{figure}

The sample was thus split into 7 bins 
of $\log$(SFR) (as estimated from the radio 
luminosity), each having between 5 and 16 objects. The median $\log$(SFR)
and Balmer decrement in each bin were then found (shown as 
asterisks in Figure~\ref{fig:balmer}). A linear fit, taking
into account the errors in both quantities, results in

\begin{equation}
\left(\frac{\rm H\alpha}{\rm H\beta}\right)_{median} = 1.29 \log ({\rm SFR}) + 5.06,
\label{balmersfr}
\end{equation}

\noindent with a correlation coefficient of 0.8. 
Keeping in mind the meaning and limitations of this correlation, 
as seen in Figure~\ref{fig:balmer}, one can now test its usefulness 
as a first correction for the effect seen in Figure~\ref{fig:sfr14sfrha}.

The departure of the observed 
Balmer decrement from the Case B value of 2.86 
\citep[e.g.,][]{Brocklehurst71}, can be related to the color 
excess for nebular emission lines, $E(B-V)_{gas}$, and extinction, 
$k(\lambda)$, by

\begin{equation}
\left( \frac{\rm H\alpha}{\rm H\beta} \right)_{\rm Case~B}  = \left( \frac{\rm H\alpha}{\rm H\beta} \right)_{\rm obs} 10^{\,0.4 \,E(B-V)_{gas}\,(k_{\alpha}-k_{\beta})}.
\label{balmer}
\end{equation}

\noindent Substituting (\ref{balmersfr}) into (\ref{balmer}) gives a
relation for the color excess as a function of SFR:

\begin{equation}
E(B-V)_{gas} = \frac{2.5}{(k_{\alpha}-k_{\beta})} \log \left( \frac{2.86}{1.29 \log ({\rm SFR}) + 5.06} \right).
\label{ebvsfr}
\end{equation}

\noindent Together with an appropriate extinction curve (the 
standard Galactic extinction curve of \citet{Cardelli89} with 
$R_{V}=3.1$, found by \citet{Calzetti01} to describe well
the reddening of the ionized gas in star-forming galaxies), 
this can then be used to correct $L_{\rm H\alpha}$, and consequently, 
SFR$_{\rm H\alpha}$, for dust obscuration:

\begin{equation}
L_{\rm H\alpha} = L_{\rm H\alpha}^{obs}~10^{\,0.4 \,E(B-V)_{gas}\,k_{\alpha}}.
\label{hacorr}
\end{equation}
where $L_{\rm H\alpha}^{obs}$ can either be the observed H$\alpha$ luminosity 
or the ``effective'' H$\alpha$ luminosity derived from an observed [O{\sc ii}]
flux. 

Equation~(\ref{ebvsfr}) gives the relation between extinction and 
the {\it intrinsic} SFR. Assuming this to be 
the value given by the radio luminosity could be a good approximation, 
but would create an artificial dependence between the corrected 
\Halpha\ SFR and the one from 1.4\,GHz. Instead, since the form 
for the SFR-dependent obscuration
is monotonically increasing, an iteration over possible values for
intrinsic SFR and the corresponding obscuration can be performed 
until
the calculated obscured SFR converges with the observed value
\citep{Hopkins01}. We note that this procedure does not take into account
any absorption of ionizing photons by dust inside 
HII regions. \citet{Charlot02}, modeling the observed spectra in non-Seyfert 
galaxies, estimate that this mechanism is responsible for the loss of 
$\sim$20\% of ionizing photons. Given the large 
uncertainty associated with this value, however, we do not 
attempt any correction, noting that its magnitude would not 
significantly affect our results.

\begin{figure}
\epsscale{1.0}
\plotone{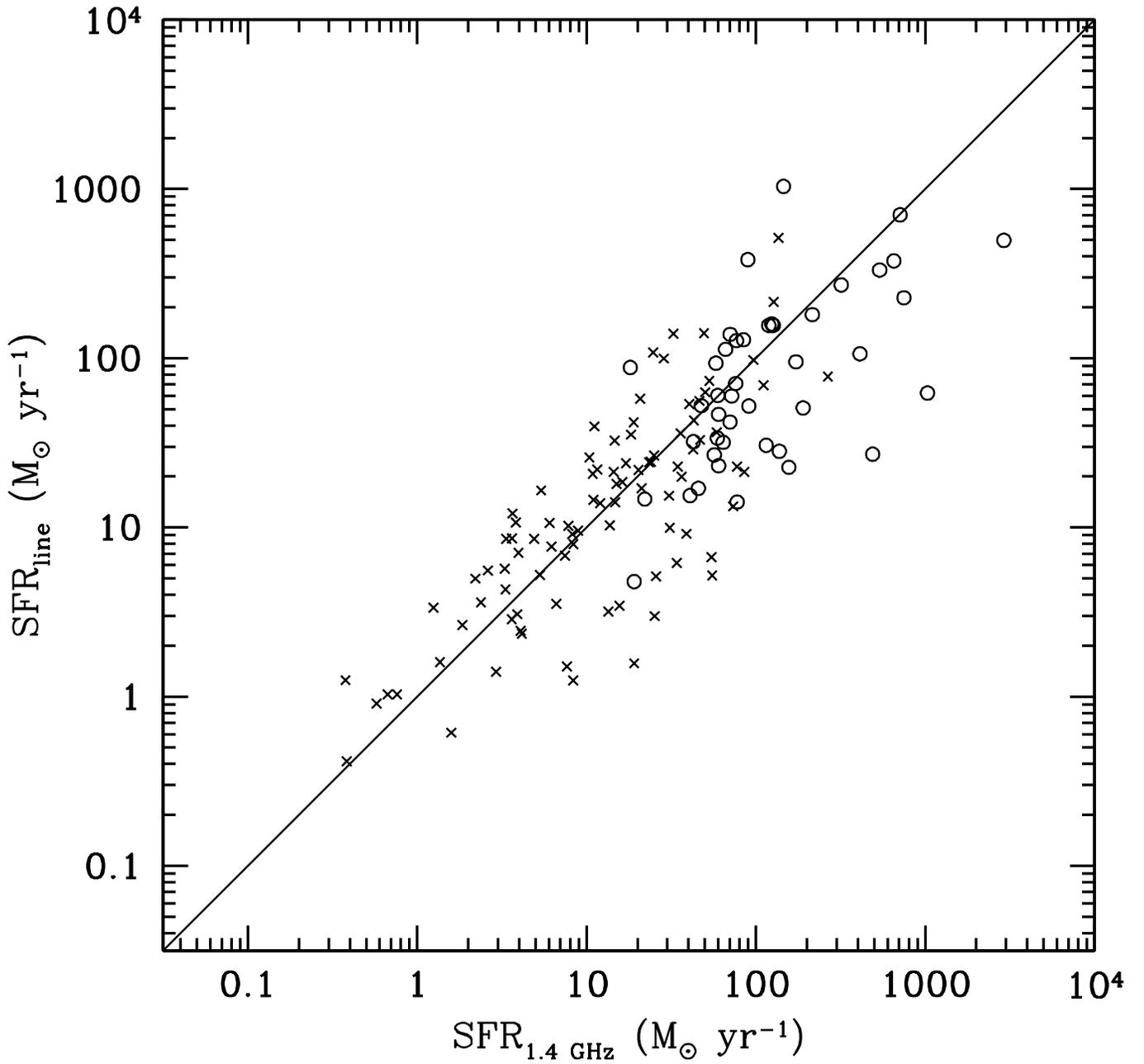}
\caption{Same as Fig.~\ref{fig:sfr14sfrha}, but with line luminosities 
corrected for dust obscuration using Equation~(\ref{hacorr}).
\label{fig:sfr14sfrhacorr}}
\end{figure}

Figure~\ref{fig:sfr14sfrhacorr} shows the 
resulting dust corrected relation for the SFR from line and 
radio luminosities. It is clear that the SFR-dependent dust 
absorption, while being a very coarse approximation, 
can successfully account for the first order offset 
between the SFRs derived from \Halpha\ or [O{\sc ii}] 
and radio luminosities for galaxies spanning a broad
range of redshifts (out to $z \approx 0.8$). This would
not be possible if the relations between Balmer decrements and SFR 
drawn from previous samples \citep{Hopkins01,Sullivan01} had been used.
The scatter still present has an rms of 0.4 dex about the best fit line,
maintained from the scatter in Figure~\ref{fig:sfr14sfrha}. The lack of
an improvement lies in the coarse relationship between
SFR and obscuration seen in Figure~\ref{fig:balmer} -- the linear fit
to the median values cannot correct for the range of obscurations seen
at each SFR. 

There will be, of course, additional uncorrelated
mechanisms involved in the \Halpha\ and radio emission which
contribute to the scatter seen, but their quantification will
only be possible after a precise account of the obscuration 
for each individual galaxy. 

\section{Summary}
A radio selected sample of star forming galaxies to
$z \approx 0.8$ has been compiled from the
{\em Phoenix Deep Survey}. The use of radio selection 
minimises bias in the sample due to dust obscuration effects. 
The relationship between obscuration and SFR is shown to be
only for a higher Balmer decrement range at higher SFRs, 
contrary to the tight correlation observed in previous studies. 
Still, the use of a linear relation, reflecting only the broadest trend, 
was explored as a first order
correction for large samples of galaxies with no direct measurement 
of obscuration. This successfully 
accounts for the major discrepancy between optical emission line SFR 
estimates and 1.4\,GHz luminosity estimates for all galaxies in the 
broad redshift range probed. However, a much more detailed correction of 
the dust 
obscuration is necessary for the study of the uncorrelated mechanisms 
(e.g., star formation histories) responsible for the scatter still present.

\acknowledgments

We thank M. Sullivan for his comments and advice. 
JA gratefully
acknowledges the support from the Science and 
Technology Foundation (FCT, Portugal) through the fellowship
BPD-5535-2001 and the research grant ESO-FNU-43805-2001. 
AMH acknowledges support provided by the National Aeronautics and Space
Administration (NASA) through Hubble Fellowship grant
HST-HF-01140.01-A awarded by the Space Telescope Science Institute (STScI).
JA dedicates this work to the
memory of Gustavo Camejo Rodrigues, who will always be
dearly remembered.

\clearpage

\end{document}